\documentstyle[prd,aps]{revtex}

\input epsf

\def\notp{p\kern-4.5pt\hbox{$/$} }
\def\u#1{$\underline{\smash{\vphantom{y}\hbox{#1}}}$}

\title{Pion Interactions in Chiral Field Theories\thanks{
     Published in {\it Mod.\ Phys.\ Lett.}\/ {\bf A14} (1999) 1349.}}
\author{M.D.~Scadron}
\address{Physics Dept. Univ. of Arizona, Tucson AZ  85721, USA}
\date{\today}

\begin{document}
\maketitle
 
\begin{abstract}
We study in various chiral models the pion charge radius, $\pi_{e3}$ form factor ratio, 
$\pi^\circ \rightarrow \gamma
\gamma$ amplitude, charge pion polarizabilities, $\gamma\gamma \rightarrow \pi^\circ
\pi^\circ$ amplitude at low energies and the $\pi\pi$ s-wave $I = 0$ scattering length.
 We find that a quark-level linear sigma approach (also being consistent
with tree-level vector meson dominance) is quite compatible with all of the above data.

\noindent
PACs numbers:  11.30.Rd,11.40.Ha,12.40Vv

\end{abstract}

\section{Introduction}

In this paper we study the interactions of observed pions with inferred scalar 
$\sigma$ meson [1,2] and fermion quark SU(2) fields in a chiral-invariant manner at
low energies.  Specifically we consider two chiral theories:

a)  A chiral quark model (CQM) dynamically inducing [2] the entire quark-level SU(2)
linear $\sigma$ model (L$\sigma$M) but depending on no free parameters.

b)  Chiral perturbation theory (ChPT) involving ten strong interaction parameters
$L_1 - L_{10}$ [3-5], now called low energy constants (LECs).

Following the surveys of Donoghue and Holstein [6,7], we compare the predictions of the
above two theories with the measured values of the i) pion charge radius, ii) $\pi_{e3}$
form factor ratio $F_A/F_V$ at zero invariant momentum transfer and the $\pi^\circ 
\rightarrow \gamma\gamma$ amplitude, iii) charged pion polarizabilities, iv) $\gamma
\gamma \rightarrow \pi^\circ \pi^\circ$ amplitude at low energies, v) $\pi\pi$ s-wave
$I=0$ scattering length.

We begin in Sec.II with the quark-level Goldberger-Treiman relation (GTR), (its meson 
analog) the KSRF relation [8] and the link to vector meson dominance (VMD).  Then in 
Sec.III we examine the pion charge radius $r_\pi$ in the above two chiral theories.  Next
in Sec IV, we first review $\pi^\circ \rightarrow \gamma\gamma$ decay and then study its 
isospin-rotated semileptonic weak analog $\pi^+ \rightarrow e^+ \nu \gamma$, giving rise
to the form factor ratio $F_A/F_V \equiv \gamma$ at zero invariant momentum transfer.  This
naturally leads in Sec.V to the charged pion electric polarizability $\alpha_{\pi^+}$
due to the model-independent relation [9,6] between $\alpha_{\pi^+}$ and the
above $\pi_{e3}$ ratio $\gamma$.  Finally in Sec.VI we review the Weinberg soft-pion 
prediction [10] for the s-wave $I = 0 \ \pi\pi$ scattering length and its chiral-breaking 
corrections.

In all of the above cases the predictions of the CQM-L$\sigma$M and ChPT chiral theories
are compared with the measured values of $r_\pi, \gamma, \alpha_{\pi^+}, a^{(0)}_{\pi\pi}$.
We review these results in Sec.VII.

\section{CQM Link to GTR, VMD and KSRF}
 
The chiral quark model (CQM) involves $u$ and $d$ quark loops coupling in a chiral invariant
manner to external pseudoscalar pions (and scalar $\sigma$ mesons).
In order to manifest the Nambu-Goldstone theorem with $m_\pi = 0$ and conserved axial
currents $\partial A^{\vec{\pi}} = 0$, it is clear that the quark-level 
Goldberger-Treiman relation (GTR) must hold:
$$
f_\pi g_{\pi qq} = m_q.
\eqno(1)
$$
Here the pion decay constant is $f_\pi \approx 90 MeV$ in the chiral limit [11] and the 
constituent quark mass is expected to be $m_q \sim m_N/3 \sim 320 MeV$.  Indeed, this dynamical
quark mass $m_q \sim 320 MeV$ also follows from nonperturbative QCD considerations [12], scaled
to the quark condensate.   

Given these nonperturbative mass scales of 90 MeV and 320 MeV, the dimensionless pion-quark
coupling should be $g_{\pi qq}\sim 320 /90 \approx 3.6$.  The latter scale of 3.6 also
follows from the phenomenological $\pi NN$ coupling constant [13] $g_{\pi NN} \approx
13.4$ since then
$$
g_{\pi qq} = g_{\pi NN}/3g_A \approx 3.5
\eqno(2)
$$
for the measured value [14] $g_A \approx 1.267$.  In fact in the SU(2) CQM with $u$ and $d$
loops for $N_c = 3$, cutoff-independent dimensional regularization dynamically generates the
entire quark-level linear sigma model (L$\sigma$M) and also requires [2]
$$
g_{\pi qq} = 2\pi/\sqrt{3} \approx 3.6276 \ \ {\rm and } \ \ m_\sigma = 2 m_q.
\eqno(3)
$$
The former coupling is compatible with (1) and (2) and the latter scalar-mass relation 
also holds in the four-quark chiral NJL scheme [15] in the chiral limit.  If one 
substitutes $g_{\pi qq} = 2\pi/\sqrt{3} $ back into the GTR (1), one finds
$$
m_q = f_\pi  2\pi/\sqrt{3} \approx 325\ MeV\ \ {\rm and} \ \ m_\sigma = 2 m_q \approx 650\ MeV.
\eqno(4)
$$

Moreover the CQM quark loop for the vacuum to pion matrix element of the axial current 
$\langle 0| \overline{q} {1 \over 2} \lambda_3 \gamma_\mu \gamma_5 q | \pi^\circ \rangle = i f_\pi q_\mu$ as
depicted in Fig.1, generates the log-divergent gap equation in the chiral-limit once the GTR
(1) is employed:
$$
1 = -i 4 N_c g^2_{\pi qq} \int^\Lambda_0 {d^4 p \over (2\pi)^4 } {1 \over (p^2-m^2_q)^2}.
\eqno(5)
$$
Given the pion-quark coupling in (2) or (3), it is easy to show that the cutoff in (5)
must be $\Lambda \approx 2.3 \ m_q \approx 750 MeV$,  This naturally separates the 
``elementary" $\sigma$ with $m_\sigma \approx 650 MeV$ in (4) from the ``bound
state" $\rho$ meson with $m_\rho \approx 770 MeV$.

In fact it was shown in the third reference in [1] that the CQM $u$ and $d$ quark
loops of Fig 2 for $\rho^\circ \rightarrow \pi^+\pi^-$ lead to the chiral-limiting
relation
$$
g_{\rho \pi\pi} = g_\rho \left[ -i 4 N_c g^2_{\pi qq} \int
{d^4 p \over (2\pi)^4 }{1 \over (p^2-m^2_q)^2}  \right]  = g_\rho,
\eqno(6)
$$
where the gap equation (5) is used.  Experimentally [14] 
$g_{\rho \pi\pi}^2 / 4\pi \approx 3.0$ or $|g_{\rho \pi\pi}| \approx 6.1$, while the rho-quark
coupling measured in $\rho^\circ \rightarrow e^+e^-$ is $|g_\rho| \approx 5.0$.  From the
perspective of vector meson dominance (VMD), equ. (6) is the well-known VMD universality
relation [16].  Moreover CQM quark loops with an external $\rho^\circ$ replaced by a 
photon $\gamma$ corresponds to the VMD $\rho^\circ - \gamma$ analogy [17].  However from
the perspective of the dynamical generated L$\sigma$M, $g_{\rho \pi\pi} = g_\rho$ in
(6) corresponds to a Z = 0 compositeness condition [18].  It shrinks ``loops
to trees", implying that the L$\sigma$M analogue equation $g_{\sigma \pi\pi} = g^\prime$ 
can treat the $\sigma$ as an elementary particle while the NJL model can treat the 
$\sigma$ as a $\bar{q}q$ bound state.

Lastly, the meson analogue of the fermion GTR (1) is the KSRF relation [8], generating the
$\rho$ mass as 
$$
m_\rho \approx \sqrt{2} f_\pi (g_\rho g_{\rho \pi\pi})^{1/2}  \approx 730 \ MeV.
\eqno(7)
$$
We recall that (7) also follows by equating the $I = 1  \ \pi N$ VMD $\rho$-dominated
amplitude $g_{\rho \pi\pi} g_\rho / m^2_\rho$ to the chiral-symmetric current algebra
amplitude $1/2 f^2_\pi$ [19].  In short, the CQM quark loops combined with the 
quark-level GTR (1) dynamically generate the entire L$\sigma$M and the NJL relation
(3), along with the VMD universality and KSRF relations (6) and (7).  This collective
CQM-L$\sigma$M-NJL-VMD-KSRF picture [20] will represent our first chiral approach 
to pion interactions as characterized by $r_\pi, F_A/F_V, \alpha_{\pi^+}$ and 
$a^{(0)}_{\pi\pi}.$

\bigskip

\section{Pion Charge Radius}

It is now well-understood [21] that the CQM quark loop-depicted in Fig 3 generates the pion charge
radius (squared) for $N_c = 3$ in the chiral limit with $f_\pi \approx 90 \ MeV$ as 
$$
\langle r^2_\pi \rangle = {3 \over 4\pi^2 f^2_\pi} \approx (0.60 fm)^2.
\eqno(8)
$$
Stated another way, using the CQM-L$\sigma$M $g_{\pi qq} = 2\pi/\sqrt{3}$ coupling relation in (3),
$r_\pi$ in (8) can be expressed in terms of the GTR as the inverse Compton mass
$$
r_\pi = {\sqrt{3} \over 2\pi f_\pi} = {1 \over g_{\pi qq} f_\pi}={1 \over m_q} \approx 0.61 \ fm,
\eqno(9)
$$
using the quark mass scale in (4).  In either case this predicted pion charge radius is quite close to the
measured value [22] of 0.63 fm. A CQM interpretation of (9) is that the quarks in a Goldstone $\bar{q}$q
pion are tightly bound and $\emph{fuse}$ together, so that $m_\pi=0$ in the chiral limit with pion charge
radius $r_\pi=1/m_q$ the size of just $\emph{one}$ quark.   

Another link of $r_\pi$ to the CQM-L$\sigma$M-VMD-KSRF picture derives from examining the standard
VMD result
$$
r_\pi = {\sqrt{6} \over m_\rho} \approx 0.63\ fm.
\eqno(10)
$$
Not only is (10) in agreement with experiment, but equating the square root of (8) to (10) and invoking
the KSRF relation (7) in turn requires with $g_{\rho \pi\pi} = g_\rho$,
$$
g_{\rho \pi\pi} = 2\pi  \approx 6.28.
\eqno(11)
$$
This relation has long been stressed in a L$\sigma$M context [23], and is of course compatible
with the measured $\rho \rightarrow 2\pi$ coupling $|g_{\rho \pi\pi}| \approx 6.1.$

But a deeper CQM-L$\sigma$M connection exists due to (11).  In ref [2] the CQM quark loops of 
Fig 4 for the vacuum to $\rho^\circ$-matrix element $\langle 0 | V^{em}_\mu | \rho^\circ \rangle$ 
$= (em^2_\rho/g_\rho) \varepsilon_\mu$ was shown to dynamically generate the vector polarization
function $\Pi(k^2,m_q)$ in the chiral limit $k^2 \rightarrow$ 0, 
$$
{1 \over g^2_\rho} = \Pi (k^2=0,m_q) = -{8iN_c \over 6} \int {d^4 p \over (2\pi)^4} {1 \over (p^2-m^2_q)^2}
= {1 \over 3g^2_{\pi qq}},
\eqno(12)
$$
by use of the gap equation (5).  Then invoking the CQM-L$\sigma$M coupling $g_{\pi qq} = 2\pi/\sqrt{3}$
from (3), equation (12) together with the VMD relation (6) leads to 
$$
g_\rho = g_{\rho \pi \pi} = \sqrt{3} g_{\pi qq} = 2\pi,
\eqno(13)
$$
which recovers (11).

The second chiral approach, referred to as chiral perturbation theory (ChPT), is not considered as a model 
but a method relating various chiral observables.  However the cornerstone of ChPT is that the pion charge
radius $r_\pi$ $\emph{diverges}$ [24] in the chiral limit (CL) and that away from the CL $r_\pi$ is fixed by
the LEC $L_9$ as
$$
\langle r^2_\pi \rangle = 12 \ L_9/f^2_\pi + {\rm chiral\ loops}.
\eqno(14)
$$
To the extent that $L_9$ is scaled to the VMD value of $r_\pi$ in (10) and the chiral
loops in (14) are small [24], this ChPT-VMD approach leads to reasonable phenomenology, as 
emphasized in ref. [6].  But from our perspective, this ChPT relation (14) circumvents the physics of
(8)-(13).  Instead the CL $r_\pi$ is $\emph{finite}$ and is 0.60-0.61 fm in (8) or (9), near the measured
value $0.63\pm 0.01$ fm.  The LEC $L_9$ does not explain this fact. 

\section{$\Pi^+_{e3}$ Form Factors and $\pi^\circ \rightarrow 2\gamma$ Decay} 

The CQM $u$ and $d$ quark loops for $\pi^\circ \rightarrow 2\gamma$ decay in Fig 5
generate the Steinberger-ABJ anomaly amplitude [25] $F_{\pi^\circ\gamma\gamma}
\varepsilon_{\mu\nu\alpha\beta} (\varepsilon^{*\mu}\varepsilon^\nu k^\prime k)$ where
$$
|F_{\pi^\circ\gamma\gamma}| = {\alpha \over \pi f_\pi} \approx 0.0258 \ GeV^{-1}
\eqno(15)
$$
in the $m_\pi = 0 $ chiral limit, using the quark-level GTR (1).  Since no pion loop can contribute
to $\pi^\circ \rightarrow 2\gamma$, the CQM-Steinberger-ABJ anomaly result (15) is also the
L$\sigma$M amplitude.  Then with $m_q \approx 325$ MeV traversing the quark loops in Fig 5, the
$\pi^\circ\gamma\gamma$ decay rate from (15) is predicted to be [26]
$$
\Gamma_{\pi^\circ\gamma\gamma} = {m^3_\pi \over 64\pi} |F_{\pi^\circ\gamma\gamma}|^2\approx
8\ eV \left[ {2m_q \over m_\pi} \sin^{-1} \left( {m_\pi \over 2m_q} 
\right) \right]^4 \approx 8\ eV
\eqno(16)
$$
with $m_\pi / 2 m_q \approx  0.21 << 1$.  Of course the latter rate in (16) is near the observed value [14] 
$(7.74 \pm 0.6)$ eV.

Treating $\pi^+ \rightarrow e^+ \nu \gamma$ as an off-shell version of $\pi^\circ \rightarrow
\gamma\gamma$ decay, the CVC SU(2) rotation of (15) predicts the zero momentum transfer vector form
factor [27]
$$
F_V (0) = {\sqrt{2}\over 8\pi^2 f_\pi} \sim 0.19 \ GeV^{-1} \sim 0.027 m^{-1}_\pi.
\eqno(17)
$$
A pure quark model is then in doubt [28], because the analogue axial vector quark loop is identical
to (17) so that $\gamma_{qk} = F_A(0)/F_V(0)|_{qk} = 1$, which is about twice the observed $\gamma$.
In fact the 1998 PDG values [14], statistically dominated by the same experiment (minimizing the
systematic errors) gives
$$
\gamma_{exp} = {F_A(0) \over F_V(0)} = {0.0116 \pm 0.0016 \over 0.017 \pm 0.008} = 0.68 \pm 0.34.
\eqno(18)
$$

However the L$\sigma$M generates both quark and meson loops to the $\pi^+ \rightarrow e^+ \nu\gamma$
amplitude as depicted in Fig 6.  This leads to the $F_A (0)$ axial current form factor [29]
$$
F_A(0) = F_A^{qk}(0) + F_A^{meson} (0) = \sqrt{2} (8\pi^2 f_\pi)^{-1} - \sqrt{2} (24\pi^2 f_\pi)^{-1}
= \sqrt{2} (12\pi^2 f_\pi)^{-1},
\eqno(19)
$$
or with a $\gamma$ found from (19) divided by (17):
$$
\gamma_{L\sigma M} = {F_A(0) \over F_V(0)} = 1 - {1 \over 3} = {2 \over 3} .
\eqno(20)
$$
It is satisfying that $\gamma_{L\sigma M}$ in (20) accurately reflects the central value of the 
observed ratio in (18).

On the other hand, the ChPT picture appears [6] to give values of $\gamma = F_A(0)/F_V(0)$
varying from 0 (in leading-log approximation) to 1 in a chiral quark model-type calculation
[6]
$$
{F_A(0) \over F_V(0)} = 32 \pi^2 (L_9 + L_{10}) = 1.
\eqno(21)
$$
The latter (incorrect) value holds when the pion charge radius is (correctly)
given by the CQM-VMD value [6]
$$
\langle r^2_\pi \rangle = {12 L_9 \over f^2_\pi } = {3 \over 4\pi^2 f^2_\pi}.
\eqno(22)
$$

\section{Charged Pion Polarizabilities and $\gamma\gamma \rightarrow \pi\pi$ scattering}

Electric and magnetic polarizabilities characterize the next-to-leading order (non-pole)
terms in a low energy expansion of the $\gamma \pi \rightarrow \gamma \pi$ amplitude.
Although in rationalized units (with $\alpha = e^2/4\pi \approx 1/137)$ the classical energy $U$
generated by electric and magnetic fields is $U = ({1\over2}) \int d^3 x (\vec{E}^2 + \vec{B}^2)$,
we follow recent convention and  define charged or neutral electric $(\alpha_\pi)$ and magnetic $(\beta_\pi)$
polarizabilities from the effective potential $V_{eff}$ as
$$
V_{eff} = - {4\pi \over 2} ( \alpha_\pi \vec{E}^2 + \beta_\pi \vec{B}^2 ).
\eqno(23)
$$
With this definition [30], $\alpha_\pi$ and $\beta_\pi$ have units of volume expressed
in terms of $10^{-4} \ fm^3 = 10^{-43} \ cm^3$.  Chiral symmetry with $m_\pi \rightarrow 0$
requires $\alpha_\pi + \beta_\pi \rightarrow 0$ for charged or neutral pion polarizabilities
and this appears to be approximately borne out by experiment.  As for the charged pion 
polarizabilities, three different experiments for $\gamma\pi^+ \rightarrow \gamma\pi^+$
respectively yield the values [31-33]
$$
\alpha_{\pi^+} = (6.8 \pm 1.4 \pm 1.2) \times 10^{-4} \ fm^3
\eqno(24)
$$
$$
\alpha_{\pi^+} = (20 \pm 12) \times 10^{-4} \ fm^3
\eqno(25)
$$
$$
\alpha_{\pi^+} = (2.2 \pm 1.6) \times 10^{-4} \ fm^3.
\eqno(26)
$$

In the CQM-L$\sigma$M scheme, the simplest way to find the charged pion electric
polarizability $\alpha_{\pi^+}$ is to link it to the $\pi_{e3}$ ratio 
$\gamma = F_A (0) / F_V (0)$ via the model-independent relation
$$
\alpha_{\pi^+} = {\alpha \over 8\pi^2 m_\pi f^2_\pi} \gamma
\eqno(27)
$$
first derived by Terent'ev [9].  Since one knows that $\gamma_{L\sigma M} = 2/3$ from
(20) (consistent with observation), the  L$\sigma$M combined with (27) predicts
$$
\alpha^{L\sigma M}_{\pi^+} = {\alpha \over 12\pi^2 m_\pi f^2_\pi} 
\approx 3.9 \times 10^{-4} \ fm^3.
\eqno(28a)
$$
This L$\sigma$M polarizability (28a) is internally consistent because a direct (but tedious)
calculation of $\alpha_{\pi^+}$ due to quark loops and meson loops gives the L$\sigma$M
value [34]
$$ 
\alpha^{L\sigma M}_{\pi^+} = \alpha^{qk}_{\pi^+} + \alpha^{meson}_{\pi^+}= 
{\alpha \over 8\pi^2 m_\pi f^2_\pi} - {\alpha \over 24\pi^2 m_\pi f^2_\pi} =
{\alpha \over 12\pi^2 m_\pi f^2_\pi} ,
\eqno(28b)
$$
in complete agreement with (28a).  This L$\sigma$M value for $\alpha_{\pi^+}$
is midway between the measurements in (24) and (26).

The recent phenomenological studies [35] of Kaloshin and Serebryakov (KS)
analyze the Mark II data [36] for $\gamma\gamma \rightarrow \pi^+ \pi^-$ and find 
$(\alpha - \beta )_{\pi^+} = (4.8 \pm 1.0) \times 10^{-4} \ fm^3$ and 
$(\alpha + \beta )_{\pi^+} = (0.22 \pm 0.06) \times 10^{-4} \ fm^3$.  These
results correspond to 
$$
\alpha^{KS}_{\pi^+} = (2.5 \pm 0.5) \times 10^{-4} \ fm^3,
\eqno(29)
$$
not too distant from the L$\sigma$M value (28) and (26), but substantially
below (24).  However (29) is very close to the ChPT prediction of Donoghue and Holstein
[6] (DH)
$$
\alpha^{ChPT}_{\pi^+} =  ({4\alpha \over m_\pi f_\pi^2})(L_9 + L_{10})
\approx 2.8 \times 10^{-4} \ fm^3,
\eqno(30)
$$
if one uses the implied value of $L_9 + L_{10}$ from $\gamma \approx 0.5$.  
But in ref.[7] they show in Figs.7 and 9 that a full dispersive calculation for
$\gamma\gamma \rightarrow \pi^+\pi^-$ (including the dominant pole term) reasonably maps out
the low energy Mark II data from $0.3 GeV < E < 0.7 \ GeV$ for any $\pi^+$ polarizability
in the range
$$
1.4 \times 10^{-4}\ fm^3 < \alpha^{DH}_{\pi^+} < 4.2 \times 10^{-4} \ fm^3.
\eqno(31)
$$
Moreover both data analyses in (24), (26) or in (29), (31) also surround the 
L$\sigma$M-Terent'ev-L'vov prediction for $\alpha_{\pi^+}$ in (28), so the
ChPT prediction (30) is not unambiguously ``gold plated" as ChPT advocates maintain.

Next we study low energy $\gamma\gamma \rightarrow \pi^\circ \pi^\circ$ scattering, where
there is no pole term and the polarizabilities $\alpha_{\pi^\circ}$ and
$\beta_{\pi^\circ}$ are much smaller than for charged pions.  Even the sign of $\alpha_{\pi^\circ}$
is not uniquely determined.  In Fig. 7a we display the comparison of the $\gamma\gamma 
\rightarrow \pi^\circ \pi^\circ$ cross section in the low energy  region $0.3 \ GeV
< E < 0.7 \ GeV$ as found from Crystal Ball data [37] and a parameter-independent dispersive
calculation (solid line) [7, 38], verses the one-loop ChPT prediction (dashed line) [39].
This graph has already been displayed in refs [5,7].  As noted by Leutwyler [5], this
first-order ``gold-plated prediction of ChPT" might cause reason to panic.  In fact,
Kaloshin and Serebryakov in their Physics Letter Fig. 1 of ref. [35], now displayed as our
Fig. 7b, show a solid line through their (gold-plated) $\gamma\gamma \rightarrow
\pi^\circ \pi^\circ$  prediction [40] made five years {\em prior} to the Crystal
Ball results.  This was based in part upon the {\em existence of a broad scalar
$\varepsilon$(700)}  i.e.  the L$\sigma$M  $\sigma$(700).  On the other hand, ChPT
theory rules out [3, 20] the existence of an $\varepsilon$(700) scalar.   Stated in reverse,
perhaps the ChPT rise of $\sigma(\gamma\gamma \rightarrow \pi^\circ \pi^\circ)$ above
10nb in the 700 MeV region (inconsistent with Crystal Ball data) could be corrected if
the $\varepsilon (700)$ (or the L$\sigma$M-NJL $\sigma$ meson in eq. (3)) were taken into
account.

To make this point in another way, recall that the decay $A_1 \rightarrow \pi(\pi \pi)_{s\ wave}$
has a very small measured rate [14] $\Gamma = (1 \pm 1)\ MeV$.  This can be understood
[41] in the context of our CQM-L$\sigma$M picture giving rise to the {\em two} quark
loop graphs in Fig 8.  Owing to the general Dirac-matrix partial fraction identity
$$
{1\over \notp - m_q} 2m_q {1\over \notp - m_q} = -\gamma_5 {1\over \notp - m_q}-
{1\over \notp - m_q}\gamma_5,
\eqno(32)
$$
there is a soft pion theorem (SPT) which forces the ``box" and ``triangle" quark 
loops in Fig 8 to interfere destructively.  Specifically the quark-level GTR in (1)
and (32) above give in the soft pion $p_\pi \rightarrow 0$ limit
$$
\langle (\pi\pi)_{sw} \pi | A_1 \rangle = \left[  - {i \over f_\pi }  \langle
\sigma_\pi | A_1 \rangle +  {i \over f_\pi } \langle\sigma_\pi | A_1 \rangle \right]
= 0 ,
\eqno(33)
$$
in agreement with the data.

Applying a similar soft-pion argument to the two neutral pion quark loop graphs in 
Fig. 9 representing the CQM-L$\sigma$M amplitude for $\gamma\gamma \rightarrow \pi^\circ 
\pi^\circ$ scattering, a quark box plus quark triangle cancellation due to the identity (32)
leads to the SPT prediction
$$
\langle\pi^\circ\pi^\circ|\gamma\gamma\rangle \rightarrow \left[ - {i \over f_\pi }
\langle \sigma | \gamma\gamma\rangle + {i \over f_\pi }
\langle \sigma | \gamma\gamma\rangle\right] = 0 .
\eqno(34)
$$
Qualitatively this ``$\sigma$ interference" may be what ref. [40] predicts and what ChPT
is lacking in the data plots of Fig. 3 and Fig. 2 in refs. [5,7] respectively, corresponding
to our Fig. 7a.

\section{$\pi\pi$ s-wave I = 0 Scattering Length}

In the context of the CQM-L$\sigma$M picture, $\pi\pi$ quark box graphs ``shrink" back
to ``tree" diagrams due to the Z = 0 [18] structure of this theory [2].  Thus one need
not go beyond the original tree-level L$\sigma$M [42,43] as recently emphasized by 
Ko and Rudaz in ref. [1].  Following Weinberg's [10] soft-pion expansion, 
Ko and Rudaz express the $\pi\pi$ scattering amplitude in ref. [1] as
$$
M_{ab,cd} = A(s,t,u) \delta_{ab} \delta_{cd} + A(t,s,u) \delta_{ac} \delta_{bd} +
A(u,t,s) \delta_{ad} \delta_{cb}
\eqno(35)
$$
and write the s channel $I = 0$ amplitude as 
$$
T^0 (s,t,u) = 3A(s,t,u) + A(t,s,u) + A(u,t,s).
\eqno(36)
$$
Then they note that the original (tree-level) L$\sigma$M predicts
$$
A(s,t,u) = -2\lambda \left[ 1 - {2\lambda f^2_\pi \over m^2_\sigma - s} \right] ,
\eqno(37)
$$
where away from the chiral limit $m_\pi \neq 0$  one knows 
$$
\lambda = {g_{\sigma\pi\pi}\over f_\pi} = {(m^2_\sigma - m^2_\pi) \over 2f^2_\pi} ,
\eqno(38)
$$
regardless of the value of $m_\sigma$ [42-44]

Substituting (38) into (37) one obtains a slight modification of the Weinberg $(s-m^2_\pi)/
f^2_\pi$ structure:
$$
A(s,t,u) = \left( {m^2_\sigma - m^2_\pi \over m^2_\sigma - s} \right)
\left( {  s- m^2_\pi \over f^2_\pi } \right),
\eqno(39)
$$
so that the s-wave I=0 scattering length at $s = 4m^2_\pi $, $t = u= 0$ becomes in the
L$\sigma$M with $\varepsilon = m^2_\pi/ m^2_\sigma \approx 0.046 $ from (3):
$$
a^{(0)}_{\pi\pi} |_{L\sigma M} \approx \left( {7 + \varepsilon \over 1 - 4
\varepsilon } \right) {m_\pi \over 32 \pi f^2_\pi } \approx (1.23) 
{ 7 m_\pi \over 32 \pi f^2_\pi } \approx 0.20 \ m^{-1}_\pi .
\eqno(40)
$$

This 23\% enhancement of the Weinberg prediction of [10] $0.16 \ m^{-1}_\pi$ is also
obtained from ChPT considerations [45].  It is interesting that ChPT simulates [46]
the 23\% enhancement found from the L$\sigma$M analysis in (37)-(40) above,
especially in light of the ``miraculous" cancellation of L$\sigma$M tree level
terms, as explicitly shown in eqs. (5.61)-(5.62) of ref. [43]  Such a L$\sigma$M-induced
cancellation instead resembles the SPT eq. (34) for $\gamma\gamma \rightarrow \pi^\circ
\pi^\circ$ where ChPT fails, whereas it simulates (good) results similar to the
L$\sigma$M for the above $a^{(0)}_{\pi\pi}$ scattering length.

To compare the L$\sigma$M prediction (40) or the similar ChPT result with data,
one recalls the $\pi\pi$ scattering length found from $K_{e4}$ decay [46]
$$
a^{(0)}_{\pi\pi} |_{exp}^{K_{e4}} = (0.27 \pm 0.04) m^{-1}_\pi ,
\eqno(41)
$$
or the $\pi\pi$ scattering length inferred from $\pi N$ partial wave data [47]
$$
a^{(0)}_{\pi\pi} |_{exp}^{\pi N} = (0.27 \pm 0.03) m^{-1}_\pi .
\eqno(42)
$$ 
On the other hand, the Weinberg-L$\sigma$M soft-pion scattering length (39) can acquire
a hard-pion correction $\Delta a^{(0)}_{\pi\pi}$ due to the resonance decay 
$f_\circ (980) \rightarrow \pi\pi$.  This was initially computed in ref. [48] based
on a $f_\circ \rightarrow \pi\pi$ decay width $\Gamma = 24 \pm 8 \ MeV$. Since this 1992
PDG decay width has increased [14] to $\Gamma = 37 \pm 7 \ MeV$, the hard-pion scattering
length correction is now (with $g^2_{f_s} = 16 \pi m^2_{f_\circ} \Gamma / 3p \approx
1.27 GeV^2$, $\xi = m^2_\pi / m^2_{f_o} \approx 0.02)$ 
$$
\Delta a^{(0)}_{\pi\pi} = {g^2_{f_o}  \over 32 \pi m_\pi m^2_{f_o} }
\left[ {5-8\xi \over 1 - 4\xi} \right] \approx 0.07 m_\pi^{-1}.
\eqno(43)
$$
Thus the entire Weinberg-L$\sigma$M hard-pion correction prediction for the 
I = 0 s-wave $\pi\pi$ scattering length from (40) and (43) is 
$$
a^{(0)}_{\pi\pi}  \approx 0.20 m_\pi^{-1} + 0.07 m_\pi^{-1} = 0.27 m_\pi^{-1}.
\eqno(44)
$$
We note that this $a^{(0)}_{\pi\pi}$ L$\sigma$M prediction (44) is in exact agreement with 
the central value of the data in (41) or (42).

\section{Summary}

In this paper we have studied low energy pion process and compared the data with the
predictions of two chiral theories: (a) the chiral quark model (CQM) and its dynamically
generated extension to the quark-level linear $\sigma$ model (L$\sigma$M); (b) modern
chiral perturbation theory (ChPT).  We began in Sec.II by showing the direct link between
the CQM-the quark-level L$\sigma$M- and the Goldberger-Treiman relation (GTR), the $Z=0$
condition and vector meson dominance (VMD), and the KSRF relation.  In Sec.III we used this
CQM-L$\sigma$M theory to compute the pion charge radius $r_\pi = \sqrt{3} / 2\pi f_\pi 
\approx 0.61\ fm$ in the chiral limit.  This agrees well with the observed [22] and VMD values
$r_\pi = \sqrt{6} / m_\rho \approx 0.63 \ fm$.  In fact setting $r_\pi^{L\sigma M} =
r_\pi^{VMD}$ leads to the rho-pion coupling $g_{\rho\pi\pi} = 2\pi$, which is only
2\% greater than the observed PDG value [14].  On the other hand, ChPT fits the measured
$r_\pi$ to the parameter $L_9$ while maintaining that chiral log corrections are small [24].

Then in Sec.IV we computed the $\pi^\circ \rightarrow \gamma\gamma$ and $\pi^+
\rightarrow e^+\nu\gamma$ amplitudes in the L$\sigma$M and found both match data with
the latter predicting $\gamma = F_A/F_V = 2/3$, while experiment gives [14]
$\gamma = 0.68 \pm 0.34$.   We extended the latter L$\sigma$M loop analysis to charged pion
polarizabilities in Sec.V, finding $\alpha_{\pi^+} = \alpha(12\pi^2 m_\pi f^2_\pi)^{-1}
\approx 3.9 \times 10^{-4}\ fm^3$, midway between the observed values.  Also we studied
$\gamma\gamma \rightarrow \pi^\circ\pi^\circ $ scattering at low energy, where data requires
an s-wave cross section $\sigma < 10 nb$ around energy $E \sim 700$ MeV, and where ChPT
predicts $ \sigma \sim 20nb$.  In contrast, refs [35,40], accounting for the 
(L$\sigma$M) scalar resonance $\varepsilon(700)$ appeared to predict $\sigma < 10 nb$
five years before the first Crystal Ball data was published [37].  Finally, in Sec VI we
extended Weinberg's soft pion (PCAC) prediction [10] for $a^{(0)}_{\pi\pi}$, the I = 0 
s-wave $\pi\pi$ scattering length, to the (tree level) L$\sigma$M.  Also, hard-pion corrections
due to the $f_o (980) \rightarrow \pi\pi$ scalar resonance decays led to an overall
scattering length $a^{(0)}_{\pi\pi} \approx 0.27 m^{-1}_\pi$ in the extended L$\sigma$M,
in perfect agreement with the central value of both the $K_{l4}$ and $\pi N$-based 
measurements [46, 47] of $a^{(0)}_{\pi\pi}$.

In all of the above cases we compared these CQM-L$\sigma$M predictions (depending upon {\em no}
arbitrary parameters) with the predictions of chiral perturbation theory (ChPT depending on 
ten parameters  $L_1 - L_{10}$) and found the latter theory 
almost always lacking.  These results were tabulated in the following Table 1.

It is important to stress that a $Z =0$ condition [18] is automatically satisfied in the
strong interaction CQM-L$\sigma $M-VMD-KSRF theory and always ``shrinks" quark loop
graphs to ``trees" for strong interaction processes such as $g_{\rho\pi\pi} = g_\rho$ or
$\pi\pi\rightarrow\pi\pi$ and its accompanying $a^{(0)}_{\pi\pi}$ scattering length.  This 
also makes VMD a tree-level phenomenology, as long stressed by Sakurai [16,19].  For
processes involving a photon, however, such as for $r_\pi$, $\pi^\circ \rightarrow 2 \gamma$,
$\pi^+ \rightarrow e^+ \nu \gamma$, $\gamma\gamma \rightarrow \pi\pi$ and for $\alpha_\pi,
\beta_\pi$, the above L$\sigma $M loop graphs must be considered (since a $Z =0$ condition
no longer applies).

In all of the above pion processes, the (internal) scalar $\sigma$ meson
plays an important role in ensuring the overall chiral symmetry (and current algebra-PCAC
in the case of KSRF, $\pi\pi\rightarrow\pi\pi$, $A_1 \rightarrow 3\pi$ and $\gamma\gamma 
\rightarrow \pi\pi$) for the relevant Feynman amplitude.  Cases in point are 
$A_1 \rightarrow \pi(\pi\pi)_{s\  wave}$, and $\gamma\gamma\rightarrow \pi^\circ\pi^\circ$,
where the internal $\sigma$ mesons in Figs. 8b and 9b ensure the soft pion theorems (SPT), 
equs. (33) and (34), which in fact are compatible with the data.  This SPT role of the
$\sigma(700)$ in $\gamma\gamma\rightarrow \pi^\circ\pi^\circ$ may explain why the ChPT approach
does not conform to Crystal Ball data in Figs. (3,5) in refs [5,7], respectively [49].  

In fact clues of a broad $\sigma(700)$ have been seen in (at least) seven  different 
experimental analyses in the past 16 years [50].  As noted in ref.[20], the ChPT 
attempt to rule out a L$\sigma$M structure was based on problems of a pure meson
L$\sigma$M (with that we agree)-but a pure meson L$\sigma$M is {\em not} the 
CQM-L$\sigma$M to which we adhere.  The latter always begins in the chiral limit with
axial current  conservation due to a {\em quark}-level Goldberger-Treiman relation (GTR)
eq.  (1), which dynamically induces the  L$\sigma$M [2] starting from (CQM)
{\em quark} loops.

As our final observation, the GTR-VMD-KSRF basis of our proposed CQM-L$\sigma$M
theory are really examples [44] of soft-pion theorems and PCAC coupled to current
algebra as used throughout the 1960's.  Staunch advocates of ChPT in ref. [51] refer to
such 1960 soft pion theorems as ``low energy guesses" [LEG].   Instead 
they prefer the strict ``low energy theorems" [LET] of modern ChPT.  However ref.  [51]
concludes with an interesting remark: ``it may be one of nature's follies that experiments 
seem to favour the original LEG over  the correct LET".  Moreover ref. [6] concludes by 
noting that the VMD approach appears to give more reliable predictions for $r_\pi$, $\gamma$,
$\alpha_\pi$ and $a^{(0)}_{\pi\pi}$ than does ChPT.  We agree with both of these statements.

The author appreciates discussions with A. Bramon, S. Coon, R. Delbourgo, V. Elias, N. Fuchs, A. Kaloshin
and R. Tarrach.

\newpage

\leftline{\bf Figure Captions}
\begin{description}
\bigskip
\item[Fig. 1]  Quark loops for axial current $\langle 0| \bar{q} {1\over 2} \lambda^3
\gamma_\mu \gamma_5 q | \pi^\circ \rangle = i f_\pi q_\mu$.
\bigskip
\item[Fig. 2]  Quark loops for $\rho^\circ \rightarrow \pi^+\pi^-$.
\bigskip
\item[Fig. 3]  Quark loops for $r_\pi$.
\bigskip
\item[Fig. 4]  Quark loops for $\rho^\circ \rightarrow \gamma$ em vector current.
\bigskip
\item[Fig. 5]  Quark loops for $\pi^\circ \rightarrow \gamma\gamma$ decay.
\bigskip
\item[Fig. 6]  Quark (a) and meson (b) loops for $\pi^+ \rightarrow e^+\nu \gamma$ in the L$\sigma$M.
\bigskip
\item[Fig. 7]  Plots of Crystal Ball $\gamma\gamma\rightarrow \pi^\circ \pi^\circ$ data verses (a) the ChPT
prediction; (b) the Kaloshin-Serebryakov prediction accounting for the $\varepsilon (700)$ scalar meson.
\bigskip
\item[Fig. 8]  Quark box (a) and triangle (b) graphs for $A_1 \rightarrow \pi(\pi\pi)_{s\ wave}$ decay.
\bigskip
\item[Fig. 9]  Quark box (a) and triangle (b) graphs for $\gamma\gamma \to \pi^\circ \pi^\circ$ scattering. 
\end{description}
 
\newpage

\leftline{\bf Table I}
\center{Comparison of chiral theory predictions of pion processes with experiment}

\bigskip
\bigskip
\bigskip

\noindent
\ \ \ \ \ \ \ \u{CQM-L$\sigma$M} \hskip 1in \u{ChPT to one loop} \hskip 1.25in \u{Data}

$$
\begin{array}{llll}
 r_\pi \ \ \ \ \ \ \ \ \ \ \ & 0.61 fm \ \ \ \ \ \ &{12\ L_9  \over f^2_\pi} + \rm chiral\ loops  & (0.63 \pm 0.01) \ fm \\
\\ \\
 g_{\rho\pi\pi}  & 2\pi \approx 6.28  &\ \ \ \   ?  & 6.1 \pm 0.1 \\ \\ \\
F_{\pi\gamma\gamma}  & \alpha / \pi f_\pi \approx 0.0258 \ GeV^{-1} &  \ \ \ \ ? & (0.026 \pm 0.001 ) \ GeV^{-1} \\ \\ \\
\gamma = {F_A(0) \over F_V(0)} & {2 \over 3} & 32\pi^2(L_9 + L_{10}) & 0.68 \pm 0.34 \\ \\ \\
\alpha_{\pi^+} & 3.9 \times 10^{-4}\ fm^3 &({4\alpha \over m_\pi f_\pi^2})(L_9 + L_{10})
 & (2.2 \ to \  6.8 ) \times 10^{-4}\ fm^3\\ \\ \\
\gamma\gamma \to \pi^\circ\pi^\circ & \sigma (E \sim 0.7\ GeV)\sim 7nb & \sigma (E\sim 0.7\ GeV)\approx 20nb &
\sigma (E \sim 0.7\ GeV) < 10nb \\ 
\ \ & \rm and \ \rm falling & \rm and  \ \rm rising & \rm and \ \rm falling \\
\\ \\ 
a^{(0)}_{\pi\pi} & 0.27 m^{-1}_\pi & 0.20 m^{-1}_\pi & (0.27 \pm 0.04) m^{-1}_\pi

\end{array}
$$

\setlength{\unitlength}{1cm}

\newpage

\begin{figure}
\begin{picture}(14,20)
\put (0.5,0.0){\epsfxsize=13cm \epsfbox{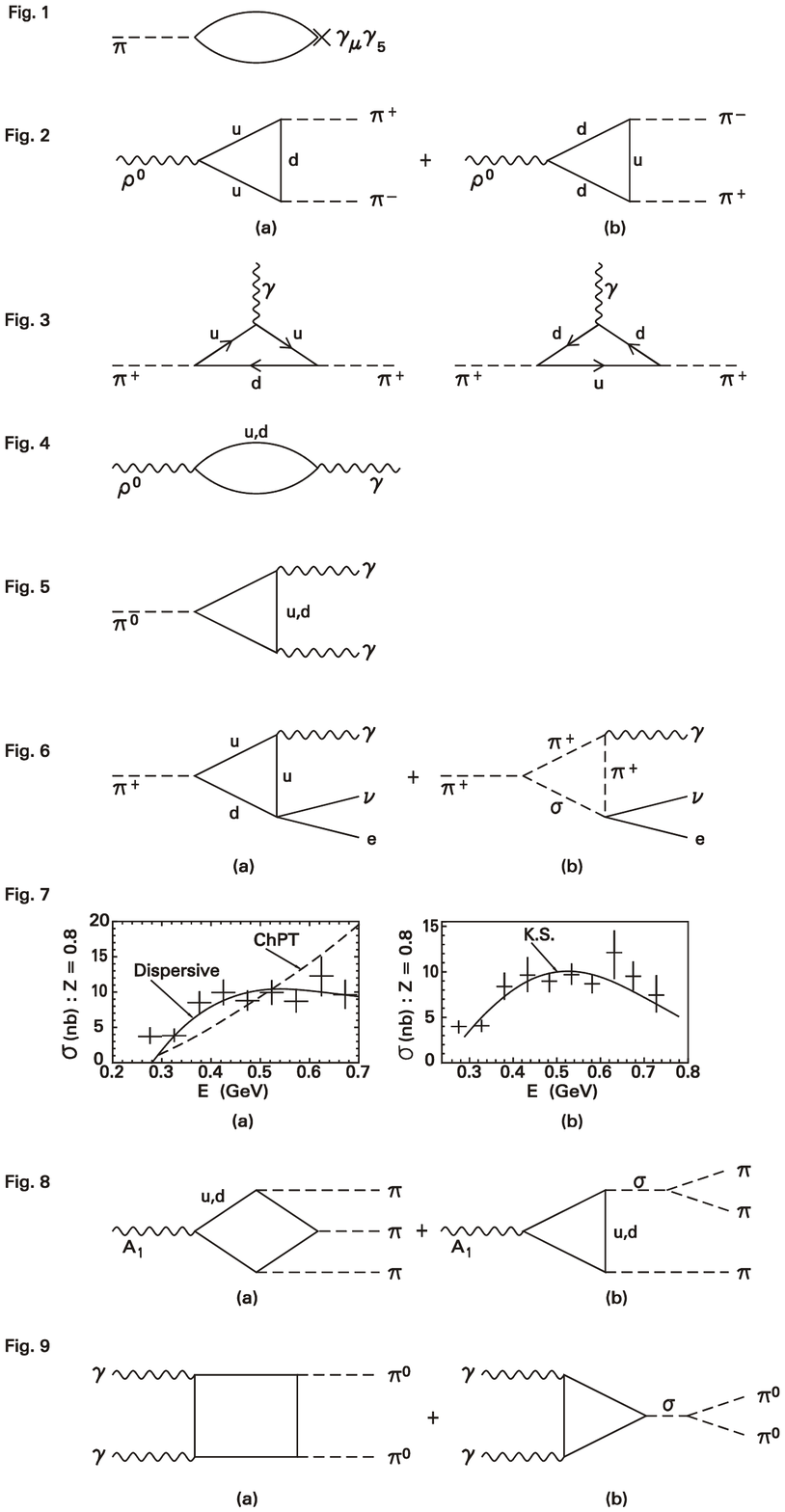}}
\end{picture}
\end{figure}


\begin{thebibliography}{99}
\bibitem{30}  An alternative (alt) definition of $\alpha_\pi, \beta_\pi$ eliminates the 4$\pi$ factor in
eq. (23), meaning that $\alpha_\pi^{alt}$ is 4$\pi$ larger than e.g. $\alpha_{\pi^+}$ in (24)-(26), then
measured in units $10^{-3}\ fm^3$.
\bibitem{44}  Also note that just as the $ I = 1\ \rho$ pole in the $q^2 \to 0$ limit $g^2_{\rho\pi\pi} / m^2_\rho$ 
simulates the current algebra-VMD-KSRF factor $1/ 2f^2_\pi$ for  $\pi\pi$ scattering, the $I=0$ $\sigma$ pole in the
L$\sigma$M quartic limit $\lambda / m^2_\sigma$ 
simulates the current algebra-Weinberg factor $1/2f^2_\pi$.
\end{thebibliography}
\end{document}